\documentstyle[aps,prl]{revtex}

\begin{document}
\draft
\title{
Prolate dominance of nuclear shape caused by
a strong interference between the effects of spin-orbit and $l^2$ terms
of the Nilsson potential
}
\author{Naoki Tajima and Norifumi Suzuki}
\address{Department of Applied Physics, Fukui
University, Bunkyo 3-9-1, Fukui, 910-8507, Japan}
\date{\today}
\maketitle
\begin{abstract}
The origin of the dominance of prolate shapes over oblate ones of the
ground states of atomic nuclei is investigated with the
Nilsson-Strutinsky method.  The number of prolate nuclei among all the
deformed even-even nuclei is calculated as a function of the strengths
of the spin-orbit and the $l^2$ terms of the Nilsson potential.  The
latter simulates a square-well like radial profile of the mean
potential.  The ratio of prolate nuclei is 86\% with the standard
strengths corresponding to the actual atomic nuclei.  By weakening the
spin-orbit potential, the ratio oscillates strongly, having a local
minimum value of 45\% with a half of the standard strength and a local
maximum value of 78\% without the spin-orbit potential.
\end{abstract}




A basic question in nuclear physics is why atomic nuclei have a strong
tendency to deform into prolate shapes than oblate ones.

Since the early days of the discovery of nuclear
deformation\cite{Rai50}, it has been usually believed that the nuclear
deformation can be ascribed to the shell structure of nucleon's
single-particle spectrum.  One might suspect the existence of some
unknown simple and direct correspondence between the prolate dominance
and a feature of the hamiltonian, e.g., a specific term of the
elementary nucleon-nucleon interaction, in an analogous fashion as the
tensor force causes a mixture of $d$ wave in the wavefunction of a
deuteron.  However, it is probably sufficient at the present stage to
confine the scope of the investigation to the mean single-particle
potential, through which most of the possible causes affect the
deformation.


There are two causes which favor prolate shapes {\em not} through the
shell effect.
One is the Coulomb repulsion between protons, which tends to deform
the nucleus into an elongated shape rather than a flattened shape.
This effect is, however, important only in heavy nuclei while the
prolate dominance is present already in middle-weight nuclei.
The other, argued by Zickendraht\cite{Zic85}, is
the difference of the volume element of
the collective coordinates between prolate and oblate
shapes, which can be identified with the difference of the
available configuration space in spherical shell model
calculations.  In mean-field approaches, this effect corresponds to
that of an angular momentum projection into spin zero states.
However, it does not seem to be essential because the prolate
dominance can be reproduced without the projection in
shell-correction\cite{MNM94} and mean-field\cite{TTO96} methods.


Let us mention three kinds of potentials which give rise to
shell effects favoring prolate shapes. 

The first one is the anisotropic harmonic oscillator, which is the
most simple approximation used for the nuclear mean-field potential.
Concerning the $sd$ shell nuclei, Bohr and Mottelson stated that
prolate (oblate) shape is preferred in the beginning (end) of the
major shell filling due to the strong shape driving effect of the
particles (holes) in the $\Omega = \frac{1}{2}$ orbital\cite{BM75}.
This seems to suggest an equal number of prolate and oblate nuclei.
According to Ref.~\cite{BM75}, it is the spin-orbit potential which
breaks this even situation by weakening the oblate-shape shell effect.
A more general argument given by Castel et al.~\cite{CRZ90} is that
the summation of the single-particle energies of an isotropic
harmonic oscillator is decreased by extending one axis and shrinking
the other two axes under volume conservation condition
neglecting detailed effects of the Pauli principle.  (Their
argument seems to apply only to the harmonic oscillator potential
contrary to their insist.)  Therefore, harmonic oscillator potentials
are expected to favor prolate shapes.  A quantitative estimation of
this effect is one of the aims of our study.

The second kind of potentials are those with square-well like radial
profile.  The nuclear mean potential resembles the Woods-Saxon
potential\cite{RS80}, which is in between a square well and a harmonic
oscillator.  Frisk found \cite{Fri90} that such radial dependence is
an origin of the prolate dominance from an analysis of classical
periodic orbitals in an ellipsoidal cavity.  By considering the volume
conservation, he showed that the shell effect at the Fermi surface
changes strongly in the prolate side while stays almost constant in the
oblate side as a function of the magnitude of deformation.
Consequently, if the spherical shape is unstable, oblate shapes are
equally unstable but there must be a more stable state in the prolate
side.

The third one is the spin-orbit potential, which is indispensable for
the reproduction of the spherical magic numbers and is an important
component of the nuclear mean potential. Its relation to the prolate
dominance is suggested from an extensive Skyrme-Hartree-Fock
calculation\cite{TTO96}: The energy difference between prolate and
oblate minima exhibits a clear and abrupt change of behavior between
$Z, N < 40$ and $Z, N > 50$ where $Z$ and $N$ are the numbers of
protons and neutrons, respectively.  In the former region prolate and
oblate solutions appear evenly in the ground state, while in the
latter region the oblate solutions have systematically higher energies
than prolate ones.  Between the two regions, the character of major
shells change from the harmonic oscillator type to the Mayer-Jenzen type,
the latter of which includes a high-$j$ intruder in each major shell
due to the spin-orbit potential.  This parallelism suggests that the
spin-orbit potential plays an essential role in giving rise to the
prolate dominance.


In this paper we will examine how the situation of the prolate
dominance changes when the radial profile of the potential and the
strength of the spin-orbit potential are different from those of
actual nuclei.  For this purpose, we employ the Nilsson-Strutinsky
method\cite{NTS69}, 
which is a convenient and well-established method to reproduce
nuclear shapes.  The single-particle potential of the method is called
the Nilsson or the modified oscillator potential and is expressed as,
\begin{eqnarray}
U(\mbox{\bf r}) &
 = & \frac{1}{2} \left( 
\omega_{\bot}^2 x^2  + \omega_{\bot}^2 y^2  +
\omega_{\|}^2 z^2  \right)
+ 2 \hbar \omega_0 r_{\rm t}^2 \sqrt{\frac{4\pi}{9}}\epsilon_4 
Y_{40}(\hat{\mbox{\bf r}})
\nonumber
\\
& & + 2 f_{ls} \kappa_N \hbar \stackrel{\circ}{\omega}_0 
\mbox{\bf l}_{\rm t} \cdot \mbox{\bf s}
- f_{ll} \kappa_N \mu_N \hbar \stackrel{\circ}{\omega}_0 \left( 
\mbox{\bf l}_{\rm t}^2 - \langle \mbox{\bf l}_{\rm t}^2 \rangle_N 
\right).
\nonumber
\end{eqnarray}
The first term stands for an anisotropic harmonic oscillator potential,
where the frequencies $\omega_{\bot}$ and $\omega_{\|}$ are expressed as
functions of a quadrupole deformation parameter $\epsilon_2$,
\begin{displaymath}
\omega_{\bot} = \omega_0 \left( 1+\frac{1}{3}\epsilon_2 \right), \;\;\;
\omega_{\|} = \omega_0 \left( 1-\frac{2}{3}\epsilon_2 \right),
\end{displaymath}
while $\omega_0$ is determined through a volume
conservation condition
$\omega_{\bot}^2 \omega_{\|} = \stackrel{\circ}{\omega}^3$.
The second term is a hexadecapole deformation potential.
The third term is a spin-orbit potential, in which orbital and spin
angular momenta are expressed as $\mbox{\bf l}$ and $\mbox{\bf s}$,
respectively.
Subscript t means the usage of the stretched coordinates.
The fourth term is called the $l^2$ term  or potential hereafter.
The Woods-Saxon type radial dependence of the potential is approximated
by this term including the square of the
orbital angular momentum.
The standard values given in Table~1 of Ref.~\cite{BR85} are used for
the parameters $\kappa_N$ and $\mu_N$,
which are dependent on the total of the oscillator quanta, $N$.
The factors $f_{ls}$ and $f_{ll}$ are introduced in this paper
to modify the standard potential, which is restored by 
putting $f_{ls}=f_{ll}=1$.
A convenient feature of the Nilsson potential for our study is that
the spin dependent and independent potentials can be changed
independently unlike in the Woods-Saxon potential or the relativistic
mean-field model\cite{Rei89}.

We have utilized a program provided by Y.~R.~Shimizu\cite{Shi97}. It
is based on the NICRA code\cite{BRA91} but is simplified for
non-rotating axially symmetric states, which makes the calculation
much faster.  The pairing correlation is active for single-particle
levels within $\pm 1.2 \hbar \omega_0$ from the Fermi level, while the
strengths of the pairing force are determined such that the 
pairing gap for smoothed level density
becomes $\bar{\Delta}$ = $ 13 A^{-1/2}$ MeV.  The
parameters of the macroscopic part\cite{MS67} are $a_{\rm s}=17.9439$
MeV, $\kappa_{\rm s}=1.7826$, and $R_{\rm c}=1.2249 A^{1/3}$ fm.  See
Ref.~\cite{BRA91} for the details of the model.

Calculations with the above model have been done in the following way.
We choose the values of reduction factors $f_{ll}$ and $f_{ls}$.  For
each combination ($f_{ll},f_{ls}$), we calculate the potential energy
surface (PES) curve versus $\epsilon_2$ ($-0.5 \le \epsilon_2 \le
0.5$, with $\epsilon_4$ optimized in $-0.16 \le \epsilon_4 \le 0.16$
for each $\epsilon_2$) for all the even-even nuclei with $8 \le Z \le
126$ and $8 \le N \le 184$ and between proton and neutron drip lines
predicted by the Bethe-Weizs{\"a}cker mass formula\cite{RS80}.  
The number of nuclei thus included is 1843.
We neglect the possibility of triaxial deformations since non-axial 
shapes are very rare for even-even nuclei\cite{TTO96}.
The reduction factors are taken from a square area $-1 \le f_{ll} \le
1$ and $-1 \le f_{ls} \le 1$.  We have taken the sampling spacing to
be $\Delta f_{ll}$ = $\Delta f_{ls}$ = 0.125, which has resulted in a
computation time of a few months with a latest personal computer.


For each PES curve, we have to label the shape of the ground state
as prolate or oblate.
One has to be very careful in generating an algorithm for this purpose.
Well-deformed nuclei usually have both prolate and oblate minima
and thus each minima can be labeled without ambiguity.
On the other hand, transitional nuclei often have several shallow minima 
in a large valley extending from oblate side to the prolate side. 
In such a situation, it is not meaningful to discuss which minima
has the lowest energy.
After examining a large number of PES curves, 
we have decided to consider only those nuclei which have both 
well-developed prolate and oblate minima.
The practical procedures we finally adopted are as follows:
(1) Draw a smeared PES curve obtained through a convolution 
with a weight function $\exp [-(\Delta \epsilon_2/0.05)^2 ]$.
(2) Separate the original (i.e., before the smearing) curve into
valleys by regarding local maxima of the smeared curve as the
``water-shed''.
(3) Taking up the minimum of the original curve in each valley,
if $\epsilon_2 < -0.05 (> 0.05)$ at the minimum and
$\epsilon_2 < 0.1 (> -0.1)$ at the right (left) end of the valley,
regard the minimum as representing an oblate (prolate) solution.
(4) If a nucleus has both oblate and prolate solutions satisfying the above
criteria and the deeper one is the oblate (prolate) one,
count the nucleus as an oblate (prolate) nucleus.
Denoting thus counted number of oblate (prolate) nuclei with $N_{\rm o}$
($N_{\rm p}$), we define the ratio of prolate nuclei as
$R_{\rm p} = N_{\rm p}/(N_{\rm p}+N_{\rm o})$.
$R_{\rm p}$ may take values from $0$ to $1$.
The denominator $N_{\rm p}+N_{\rm o}$ is about 900 on the average.
Note that the smeared curve is used only to divide the curve into valleys
and it does not affect the energy or the deformation of the minima.


Fig.~\ref{fig:Rp_13_H} shows the ratio of prolate nuclei $R_{\rm p}$
as a function of the reduction factors ($f_{ll}, f_{ls}$) by means of
contours for $R_{\rm p}$ and symbols for the locations of local maxima
(triangles) and minima (squares).  Fourth order polynomials in $f_{ll}$
or $f_{ls}$ are used for the interpolations to draw the contours and
locate the extrema.

The standard nuclear potential corresponds to top-right corner
of the figure, where $R_{\rm p}$ takes the largest value in the entire
square area.  The value is $R_{\rm p}=86\%$: Among about 900 even-even
nuclei having both prolate and oblate minima, 86\% are prolate in the
ground state.  One can say that the prolate dominance is
reproduced with the standard Nilsson potential.

Our result is also in qualitative agreement with a calculation for
metallic clusters\cite{BL93}, in which prolate ground states are found
to be roughly twice as many as oblate ones in the framework of the
jellium model with infinite square well potential.
The corresponding region in Fig.~\ref{fig:Rp_13_H} is
$f_{ls}=0$ and $f_{ll} \sim 1$, where $R_{\rm p} \sim $70 - 80\%.

The minimum value of $R_{\rm p}$ is obtained at $(f_{ll}, f_{ls})$ =
$(-1,-0.125)$, where $R_{\rm p} = 40\%$, i.e., 60\% of the deformed
nuclei are oblate.  The increasing trend of $R_{\rm p}$ as a function of
$f_{ll}$ along $f_{ls}=0$ line implies that the attractive (repulsive)
$l^2$ term favors prolate (oblate) shapes.
This result confirms the view of Frisk.

On the other hand, the spin-orbit term cannot be regarded as
favoring either prolate or oblate shapes, because $R_{\rm p}$
behaves roughly symmetrically between positive and negative values of
$f_{ls}$.
The most conspicuous fact concerning the spin-orbit term found in
our study is a very strong interference with the $l^2$ term.  
In Fig.~\ref{fig:Rp_13_H}, by moving down from the top-right corner
along a line $f_{ll}=1$, $R_{\rm p}$ takes on 86, 45, 78, 44, and 81\%
for $f_{ls}$=1, $\frac{1}{2}$, 0, $-\frac{1}{2}$, and $-1$,
respectively.  
One can see two periods of oscillations in $-1 \le f_{ls} \le 1$.
Weakening the spin-orbit term by 50\% moves the ratio $R_{\rm p}$ from the
highest peak at $f_{ls}=1$ to the bottom of a deep valley at
$f_{ls}=\frac{1}{2}$, where there are more oblate nuclei than prolate
ones.  A complete disappearance of the spin-orbit term moves the
ratio to another high peak at $f_{ls}=0$ and recovers the prolate
dominance.  Combination of the two terms produces a situation
which is beyond expectation from the independent effects of each
term.


A prolate dominance as high as 80\% is realized only for restricted
combinations of the strengths of the two terms.  It may not be a
mere coincidence that the potential of actual nuclei matches one of
such rare combinations.
The same kind of subtle balance between the two terms has been
discussed concerning the pseudo-spin
symmetry\cite{HA69,AHS69,DNS87,DBB96,DBB97a,DBB97b,Gin97,MSY98}, which
holds when $\mu_N$ of the Nilsson potential is $\frac{1}{2}$ while the
standard values of $\mu_N$ are between 0.5 and 0.6.

The prolate dominance can be related to the pseudo-spin symmetry
by extending Frisk's view to particles with spin:
An attractive $l^2$ term can cause prolate dominance
{\em if the spin is decoupled from the orbital motion}.
The prolate dominance occurs at $f_{ls} = \pm 1, 0$ and $f_{ll}=1$.
The real spin is decoupled at $f_{ls}=1$ 
while the pseudo-spin is decoupled at $f_{ls}=0$.
The point at $(f_{ll},f_{ls})=(1,-1)$ might correspond to
a similar situation in which another kind of spin-like quantity
is decoupled.
There are opinions that the pseudo-spin symmetry has a relativistic
origin.  If they would be approved finally, the prolate dominance of
nuclear deformation might also be related with some relativistic aspect
of the atomic nuclei.


Let us mention some other results from our calculations
before concluding the paper.
(1) The inclusion of optimized hexadecapole deformation has a tendency
to favor prolate shapes.  By setting $\epsilon_4 = 0$, $R_{\rm p}$ is
reduced from 86\% to 82\% at ($f_{ll},f_{ls}$)=(1,1).  An average of
$R_{\rm p}$ over all the combinations of the reduction factors is
decreased from 60\% to 55\%.\\
(2) An almost pure harmonic oscillator potential (i.e., with
$f_{ls}=f_{ll}=\epsilon_4 = 0$ and weakened pairing) produces 
$R_{\rm p}$ = 55\%. 
This is an quantitative estimation of the tendency of prolate
preference predicted by Castel et al.\cite{CRZ90}


In summary, a strong interference is found between the effects of the
spin-orbit and the $l^2$ terms of the Nilsson potential.
The ratio of prolate nuclei among well-deformed even-even nuclei is 
more than 80\% by using the standard strengths for the two terms.
Multiplication of $\pm 1$ or 0 to the strength of the spin-orbit term
does not change the situation of prolate dominance. 
On the other hand, when the strength is multiplied by $\pm \frac{1}{2}$,
the ratio is less than 50\%, i.e., there are more number of oblate 
nuclei than prolate ones.
The emergence of prolate dominance for restricted combinations of the
strengths of the two terms is in a parallel situation with the 
decoupling of real or pseudo spins from the orbital motion and can be
understood by extending Frisk's view.


We are planing to study possible changes due to (1) reductions of
pairing force strengths and (2) a replacement of the Nilsson potential
with the Woods-Saxon potential.  It is also an interesting question to
which region of Fig.~\ref{fig:Rp_13_H} neutron-rich unstable nuclei,
which are waiting for experimental studies, correspond.  In such
nuclei, both terms are expected to be more or less weakened compared
with those for stable nuclei\cite{DHN94}.


One of the authors (N.T.) thanks the Department of Energy's Institute
for Nuclear Theory at the University of Washington for its hospitality
during the completion of this work.  
He also thanks Prof.\ J.P.  Draayer and Prof.\ J. Dobaczewski for
valuable discussions. The authors are grateful to Prof.\ Y.R. Shimizu
for providing a computer program to perform the Nilsson-Strutinsky
method calculation and kindly instructing its usage.


%
%
\begin{figure}
\caption{
The ratio of prolate nuclei $R_{\rm p}$. 
The abscissa and the ordinate are the reduction factor of the
strength of the $l^2$ potential ($f_{ll}$) and that of the spin-orbit
potential ($f_{ls}$) relative to the standard values.
Contours are for $R_{\rm p}=45, 50, 55, \cdots 80\%$.
Thick curves are for $R_{\rm p}$=50\%. Solid triangles
(squares) indicate the location of local maxima (minima).
}

\label{fig:Rp_13_H}
\end{figure}


%
%


\begin{references}
\bibitem{Rai50}
  J.~Rainwater, Phys.\ Rev.\ {\bf 79}, 432 (1959).
\bibitem{Zic85} 
  W. Zickendraht, Phys.\ Rev.\ Lett.\ {\bf 54}, 1906 (1985).
\bibitem{MNM94}  
  P.~M{\"o}ller, J.R.~Nix, W.D.~Myers, and W.J.~Swiatecki,
  Atomic Data Nucl. Data Tables {\bf 59}, 185 (1995).
\bibitem{TTO96} 
  N. Tajima, S. Takahara, and N. Onishi,
  Nucl.\ Phys.\ {\bf A603}, 23 (1996).
\bibitem{BM75} 
  A. Bohr and B.R. Mottelson, {\em Nuclear Structure},
  vol.\ 2 \, p.~285, (Benjamin, New York, 1975).
\bibitem{CRZ90} 
  B. Castel, D.J. Rowe, and L. Zamick, Phys.\ Lett.\ {\bf B236}, 121 (1990).
\bibitem{RS80}
  P. Ring and P. Schuck, The nuclear many-body problem
  (Springer, New York, 1980).
\bibitem{Fri90} 
  H. Frisk, Nucl.\ Phys.\ {\bf A511}, 309 (1990).
\bibitem{NTS69} 
  S.G.~Nilsson, C.F.~Tsang, A.~Sobiczewski, Z.~Szyma\'{n}ski,
  S.~Wycech, C.~Gustafson, I.~Lamm, P.~M\"{o}ller and B.~Nilsson,
  Nucl.Phys. {\bf A131}, 1 (1969).
\bibitem{BR85} 
  T. Bengtsson and I. Ragnarsson, Nucl. Phys. {\bf A436},14 (1985).
\bibitem{Rei89} 
  P.-G. Reinhard, Rep.\ Prog.\ Phys.\ {\bf 52}, 439 (1989).
\bibitem{Shi97} 
  Y.R. Shimizu, private communication.
\bibitem{BRA91} 
  T.~Bengtsson, I.~Ragnarsson and S.~{\AA}berg,
  in ``Computational Nuclear Physics 1'',
  ed. K.~Langanke, J.A.~Maruhn and S.E.~Koonin,
  p.~51 (Springer-Verlag, Berlin, 1991).
\bibitem{MS67}
  W.D.~Myers and W.J.~Swiatecki, Ark.\ Phys.\ {\bf 36}, 343 (1967).
\bibitem{BL93} 
  A.\ Bulgac and C.\ Lewenkopf, Phys.\ Rev.\ Lett.\  {\bf 71}, 4130 (1993).
\bibitem{HA69} 
  K.T.\ Hecht and A.\ Adler, Nucl.\ Phys.\ {\bf A137}, 129 (1969).
\bibitem{AHS69} 
  A.\ Arima, M.\ Harvey and K.\ Shimizu,
  Phys.\ Lett.\ {\bf B30}, 517 (1969).
\bibitem{DNS87}
  J.\ Dudek, W.\ Nazarewicz, Z.\ Szymanski, and  G.A.\ Leander,
  Phys.\ Rev.\ Lett.\ {\bf 59}, 1405 (1987).
\bibitem{DBB96} 
  J.P.\ Draayer, A.L.\ Blokhin, and T.\ Beuschel,
  Revista Mexicana de Fisica {\bf 42}, Suplemento {\bf 1}, 21-34 (1996)
  (XIX Symposium on Nuclear Physics, Proceedings, January 3-6, 1996,
  Oaxtepec, Morelos, Mexico).
\bibitem{DBB97a} 
  J.P.\ Draayer, A.L.\ Blokhin, T.\ Beuschel, and C.\ Bahri,
  Nucl.\ Phys.\ {\bf A612}, 163 (1997).
\bibitem{DBB97b} 
  J.P.\ Draayer, T.\ Beuschel, and A.L.\ Blokhin,
  Nucl.\ Phys.\ {\bf A619}, 119 (1997).
\bibitem{Gin97} 
  J.\ Ginocchio, Phys.\ Rev.\ Lett.\ {\bf 78}, 436 (1997).
\bibitem{MSY98}
  J.\ Meng, K.\ Sugawara-Tanabe, S.\ Yamaji, P.\ Ring, and A.\ Arima,
  Phys.\ Rev.\ {\bf C58}, R628 (1998).
\bibitem{DHN94} 
  J.\ Dobaczewski, I.\ Hamamoto, W.\ Nazarewicz, and
  J.A.\ Sheikh, Phys.\ Rev.\ Lett.\ {\bf 72} (1994) 981.
\end{references}
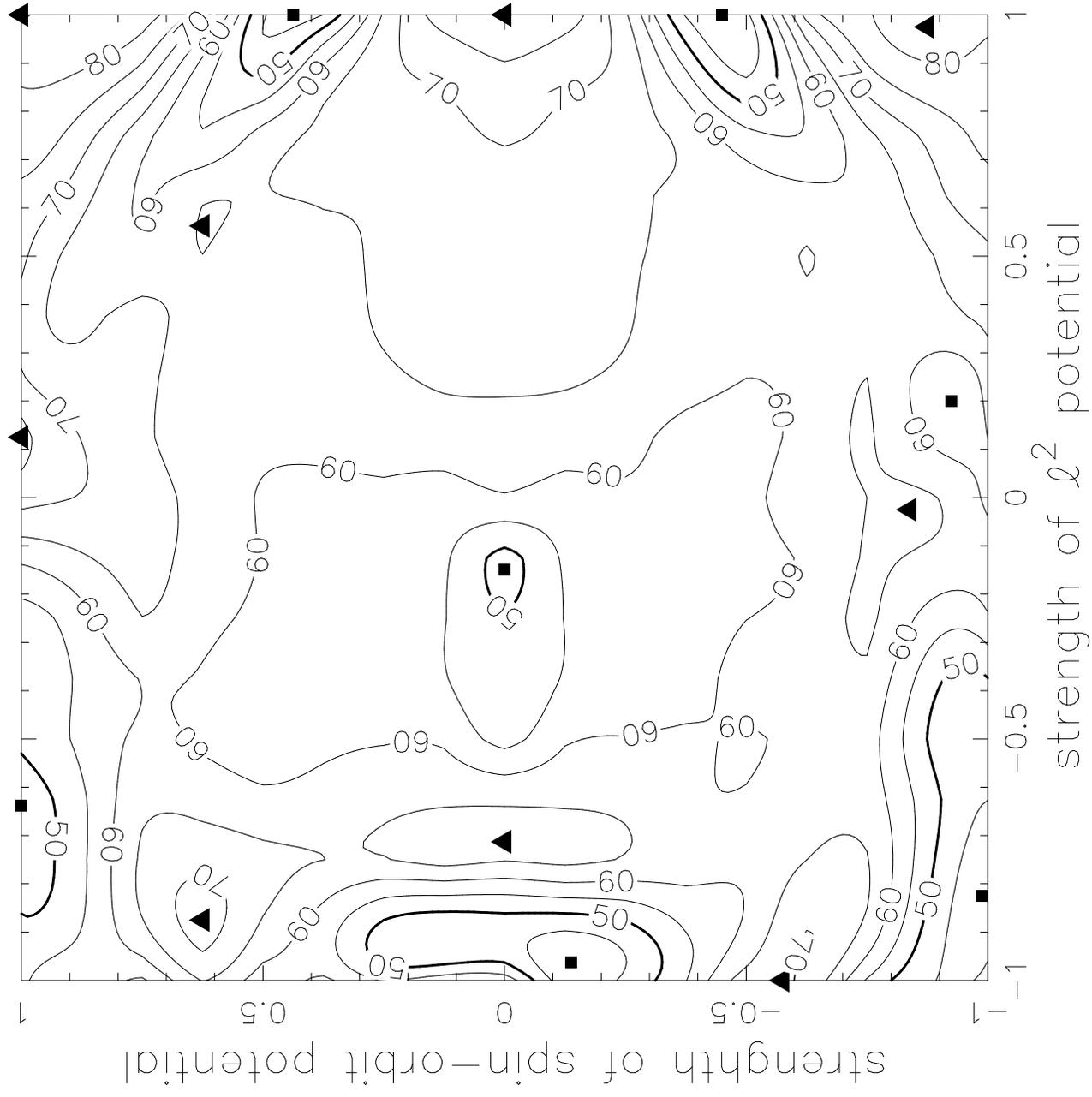
\end{document}